\documentstyle[aps,preprint]{revtex}
\begin{document}
\draft
\title{Long-Range Spatial Correlations of Eigenfunctions in Quantum Disordered Systems}

\author{V.~N.
  Prigodin$^{1,2}$, and B.~ L. Altshuler$^3$}

\address{$^1$Max-Planck-Institute f\"ur Physik komplexer Systeme,
  Heisenbergstr. 1, 70569 Stuttgart, Germany\\ $^2$A.F. Ioffe
  Physico-Technical Institute, 194021 St. Petersburg, Russia\\$^3$NEC
  Research Institute, 4 Independence Way, Princeton, NJ 08540}

\maketitle

\begin{abstract}
This paper is devoted to the statistics of the quantum eigenfunctions
in an ensemble of finite disordered systems (metallic grains). We
focus on moments of inverse participation ratio. In the universal
random matrix limit that corresponds to the infinite conductance of
the grains, these moments are self-averaging quantities. At large but
finite conductance the moments do fluctuate due to the long range
correlations in the eigenfunctions. We evaluate the distributions of
fluctuations at given conductance and geometry of the grains and
express them through the spectrum of the diffusion operator in the
grain.

\end{abstract}


\narrowtext
Weakly disordered metallic grains make an excellent laboratory to
study the phenomenon of Quantum Chaos (for a general discussion see
e.g. Ref.~\cite{AS}).  Provided electrons within a grain interact
weakly, one can describe properties of this system through one
particle quantum spectrum and corresponding eigenfunctions. The
problem is reduced to a Shr\"{o}dinger equation for a particle subject
to a potential that consists of two components: a regular potential
that confines electrons within the grain and some random potential due
to disorder.  Given the distribution of the random potential we get an
ensemble of disordered grains and can consider various statistics of
the spectra and eigenfunctions.

Classical motion in a random potential is diffusive (provided the
grain size $R$ exceeds the mean free path) with a diffusion
constant $D$.  Ensembles of weakly disordered metallic grains can
be characterized by the ``dimensionless conductance''  $g$
determined as a ratio $g=t_H/t_T$ of Heisenberg time
$t_H=\hbar/\Delta$ and Thouless time of the diffusion through the
grain $t_T = R^2/D$, where $\Delta$ is the mean energy level
spacing.  A grain would be called weakly disordered provided $g
\gg 1$.

The spectral statistics for the ensembles of grains in the limit $g \to
\infty$ are proven ~\cite{Efetov,Weidenmuller} to coincide with those for the
corresponding ensembles of random matrices~\cite{Mehta}.  These statistics
would be called {\it universal}.  The field theoretical way of evaluating
statistics in ensembles of disordered grains is based on supersymmetric
$\sigma$-model ~\cite{Efetov}.  When $g \to \infty$ the $0$-dimensional
$\sigma$-model can be used for straightforward evaluation of universal
statistics of both spectra and eigenfunction~\cite{Prigodin96}.

Finite $g$ corrections to the universal properties of quantum systems
recently attracted substantial interest.  The smooth part of the
spectral correlation function was evaluated in Ref.~\cite{ASh}.  The
first order in $1/g$ correction to the spectral correlation function
was evaluated by Kravtsov and Mirlin~\cite{Kravtsov}.  Nonperturbative
analysis of the problem~\cite{Andreev95} pointed out a qualitative
change of the behavior at finite $g$ - washing out the oscillations in
the two-point correlation function, and, hence, smoothing of the
singularities~\cite{Prigodin94} in the quantum dynamics at times close
to $t_H$.  A remarkable feature of the spectral statistics at finite
$g$ is that both their smooth~\cite{ASh} and their oscillating
parts~\cite{Andreev95} can be presented through the spectral
determinant of the classical diffusion operator $D\nabla^2$.  This
paper is devoted to the connection of this spectral determinant to the
long range correlations in the eigenfunctions.

In this paper we concentrate on the moments~\cite{Wegner80}
\begin{eqnarray}
I_\alpha (n)=V^{n-1}\int |\psi_\alpha ({\bf r})|^{2n}d{\bf r}~,
\label{def:In}
\end{eqnarray}
where $\psi_\alpha({\bf r})$ is an eigenfunction of the system
which corresponds to an eigenenergy $\epsilon_\alpha$, and $V$ is
the volume of the grain (for a d-dimensional cube $V = L^d$). The
$n=2$ moment, known as the inverse participation ratio, is
related to the level-velocity distribution~\cite{Mirlin95} or to
Hubbard-like interaction of two particles on the same quantum sate.

In the universal regime $g \to \infty$ each wave function is
extended over the whole volume, however only very short range
correlations persist: $\psi_\alpha({\bf r_1})$ and
$\psi_\alpha({\bf r_2})$ are correlated provided $r = |{\bf r}_1
- {\bf r}_2|$ is not much bigger than the particle
wavelength~\cite{Prigodin96,Mucciolo95,Alhassid,Srednicki}.  As a
result, the integration in Eq.(\ref{def:In}) provides
self-averaging, and $I_\alpha(n)$ do not fluctuate in the
universal regime!  They coincide with the moments $b_n$ of the
Porter-Thomas distribution~\cite{Mehta} for the intensity
fluctuations $|\psi_\alpha({\bf r})|^2$.  For unitary $(u)$ and
orthogonal $(o)$ symmetries they are equal to $I_\alpha (n) = b_n$, where
\begin{eqnarray}
b_{n}^{(u)} =  \Gamma (n+1),~~~b_{n}^{(o)} = 2^n\Gamma (n+1/2)/ \Gamma (1/2).
\label{moments:un}
\end{eqnarray}

The most striking difference of finite $g$ case from the
universal situation is the existence of spatial correlations of
wave function density even at $r$ comparable with the size of the
system.  As a result $I_\alpha(n)$ demonstrates finite
fluctuations from state to state and from sample to sample~\cite{Mirlin95}.
$I_\alpha(n)$ are characterized by their {\it distribution
functions}.  Let us consider the distribution $P_n(u)$ of
relative deviations of $I_\alpha(n)$ from $b_n$:
\begin{eqnarray}
u_\alpha (n) =  I_\alpha (n)/b_n - 1,~~P_n(u)=<\delta (u - u_\alpha(n))>,
\label{def:Pn}
\end{eqnarray}
where $<...>$ stands for ensemble averaging.

As we show below the Laplace transforms ${\tilde{P}}_{n}(s)= <\exp(-
su_\alpha(n))>$ of the distribution $P_n(u)$ for large but finite $g$
can be written as
\begin{eqnarray}
{\tilde {P}}_n(s) = \prod_{\mu\neq 0} \left[1+(n^2-n){2\Delta s \over
\pi\beta\omega_\mu}\right]^{-1/2}\nonumber\\ \equiv \sqrt{{\tilde Z}\left((n
-n^2){2\Delta s \over \pi\beta}\right)},
\label{result:ps}
\end{eqnarray}
where $\beta$ is different for T-invariant and not T-invariant
systems: $\beta^{(o)}=1$, $\beta^{(u)}=2$ (we are not considering
a simplectic case here).  $\omega_\mu$ is the spectrum of the diffusion
equation with the Neumann boundary conditions on the grain boundary
$B$:
\begin{eqnarray}
D\nabla^2\phi_\mu({\bf r}) =- \omega_\mu \phi_\mu({\bf r}),
~~~~~\nabla \phi|_B = 0.
\label{problem}
\end{eqnarray}
$\omega_\mu$ are not universal: this spectrum is determined by both $g$ and
the shape of the grain.  (We take $g =\beta \omega_1 / (2\Delta)$ as a
definition of $g$, since this ratio is proportional to the dimensionless
conductance for a rectangular grain).  However, the ground state of the
problem is spatially uniform, and corresponds to $\omega_0 = 0$.  All
universal statistics become applicable to disordered grains when $\omega_\mu
\to \infty$ for all $\mu$ except $\mu = 0$.  In this limit, known as zero mode
approximation, $P_n(u) = \delta (u)$ since Eq.(\ref{result:ps}) gives ${\tilde
  {P}}_{n}(s) = const$.  Therefore $I_\alpha(n)$ does not fluctuate in the
universal regime.

In Eq. (\ref{result:ps}) we introduced the function
\begin{eqnarray}
1/{\tilde {Z}}(z) = \prod_{\mu \neq 0}\left[1 - z/ \omega_\mu\right] =
z/ Z(z),
\label{zeta}
\end{eqnarray}
where $Z(z)$ is the dynamical Ruelle zeta function~\cite{Ruelle},
associated with the diffusion operator in Eq. (\ref{problem}). As it
is shown in Ref.~\cite{AAA} the pair spectral correlation
function for an ensemble of disordered grains also can be expressed
through ${\tilde Z}(z)$-function. However the spectral statistics are
determined by $|{\tilde Z}(iz)|^2$, while for the distributions
$P_{n}(u)$ one has to evaluate $Z(z)$ at real negative $z$ or to
determine both modulus and phase of $Z(iz)$.  An interesting feature
of the distributions $P_{n}(u)$ is that, in contrast with other
statistics of the quantum eigenfunctions, they are independent on the
diffusion operator eigenfunctions $\phi_\mu $, and are determined
solely by the spectrum of Eq.~\ref{problem}.

Let us describe the main features of the distributions
$P_{n}(u)$.  It is clear that long scale correlations increase
the mean value $<I(n)> = b_n [1+<u>]$ of $I_\alpha (n)$ as
compared with Eq.(\ref{moments:un}).  According to
Eq.(\ref{result:ps})
\begin{eqnarray}
P_{n}(u) = {1 \over (n^2 -n)}P(w),~~~~~~~ w = u_n /(n^2 - n),
\label{meanw}
\end{eqnarray}
and for $<u> = \int u P_{n}(u) du$ we obtain
\begin{eqnarray}
{<u> \over {n^{2} - n}} = <w> =
{-1 \over {n^2 - n}}{d \over ds}{\tilde {P}}_n(s)|_{s=0} =
{\Delta \over {\beta\pi}}\sum _{\mu \neq 0} {1 \over \omega_\mu }.
\nonumber
\end{eqnarray}
In terms of the ${\tilde Z}(z)$-function $<w>$ can be rewritten as
\begin{eqnarray}
<w> = {1 \over2}{\cal G}_{1}(0), ~~~{\cal G}_{m}(z) \equiv \left({2\Delta
\over \pi\beta}\right)^m {d^m\ln{{\tilde Z} (z)} \over dz^m}.
\label{logzeta}
\end{eqnarray}
Behavior of $P(w)$ at small $w$ can be evaluated by making a
saddle-point approximation in the inverse Laplace transformation
of Eq.(\ref{result:ps})
\begin{eqnarray}
P(w) = {\tilde Z}^{1/2}(z_c)[\pi {\cal G}_{2}(z_c)]^{-1/2}
\exp{\left[-{\pi\beta \over 2\Delta}z_c w\right]},
\label{asympsmall}
\end{eqnarray}  
\noindent provided $z_{c}(w)$ determined by the equation 
${\cal G}_{1}(z_c) = 2w$ is large, $|z_c| \gg \omega_1$. 

It follows from Eq.(\ref{result:ps}) that the probability for $w$
to be much bigger than $<w>$ is exponentially small:
\begin{eqnarray}
&P&(w) = C ~\sqrt{g /(4 w)}~\exp{\left[-\pi gw\right]},
\label{def:C}\\
C &=& \prod_{\mu\neq 0,1}~ \sqrt{1 - \omega_1/\omega_\mu}=
\left[{\tilde {Z}}(z)(1- z/\omega_1)\right]^{-1/2}\vert_{z \to \omega_1}.\nonumber
\end{eqnarray}

Consider a disordered two dimensional grain with a
particle mean free path $l$. From Eq.(\ref{problem}) it follows
\begin{eqnarray}
{\cal G}_{1}(z) = {1 \over g} \ln{\left[1 + {\omega_1 \over \omega_1 -
      z} {R^2 \over l^2}\right]},~~<w> = {\ln(R/l) \over g}.
\label{mirlin}
\end{eqnarray}
Therefore in the weak localization regime~\cite{L.R.} when all quantum states
are extended $<w> \ll 1$.  Eq. (\ref{mirlin}) for $<w>$ is in agreement with
the perturbation theory calculation~\cite{Mirlin95}.  For $ 1/g \ll w \ll <w>$
Eq. (\ref{asympsmall}) gives
\begin{eqnarray}
  P(w) = {g \over 2} \exp{\left[g(<w>-w) - {\pi \over
      2}e^{2g(<w>-w)}\right]},\nonumber
\end{eqnarray}
where $<w>$ is determined by Eq.({\ref{mirlin}).  For $ g w \lesssim 1$
\begin{eqnarray}
  P(w) = {g^{3/2} \over \sqrt{2w}} \exp{\left[g<w> -
    {\pi \over 4 g w}e^{2 g<w>}\right]}.\nonumber
\end{eqnarray}
According to Eq.(\ref{def:C}) $\ln{C} = \pi g <w> $, and at $<w> \lesssim w
\ll 1$ the distribution is
\begin{eqnarray}
  P(w) = \sqrt{g /(4w)} \exp{\left[ -\pi g (w - <w>)\right]}.
\label{morew}
\end{eqnarray}

It should be mentioned that Eqs.(\ref{result:ps}-\ref
{morew}) are valid only for $u \ll 1$.  According to
Eq.(\ref{meanw}) this means that Eqs.(\ref{result:ps} - \ref
{morew}) describe the main body of the distribution $P_n(u)$
which never takes a gaussian form.

When $u \gg 1$, the distributions $P_n(u)$ are determined by both
$\omega_\mu$ and $\phi_\mu$. This asymptotic of $P_n(u)$ can be found
by the method of optimal fluctuation.  For a spherical grain of
arbitrary dimension $d$ and radius $R$
\begin{eqnarray}
P_n(u \gg 1) \approx \exp[- a_ngu^{1/(n-1)}],
\label{result:s.p.}
\end{eqnarray}
where $g = \pi \beta D/(4R^2\Delta)$  and
\begin{eqnarray}
  a_n={d^2 \over \pi n(n-1)}\Gamma^2({1 \over 2n-2})\Gamma^{-2}({n \over
    2n-2})n^{1/(n-1)}.\nonumber
\end{eqnarray}
This distribution is valid as long as $u \ll (a_nR/l)^{n-1}$. At large
$u$ using the $\sigma$-model approach fails~\cite{SA}.

In order to derive Eq.(\ref{result:ps}) we consider mutual
distribution of $v_1 =V |\psi_\alpha ({\bf r_1})|^2$ and $v_2 = V
|\psi_\alpha ({\bf r_2})|^2$ - densities of a quantum eigenfunction
$\psi_{\alpha}(r)$.  This distribution can be reconstructed through
its moments $M_{pq} = V^{p+q} <|\psi_\alpha ({\bf r_1})|^{2p}
|\psi_\alpha ({\bf r_2})|^{2q} >$.  The latter can be calculated for a
disordered grain by analyzing moments of one-electron Green functions
using the supersymmetric $\sigma$-model technique in a way similar to
how one-point moments $M_{p0}$ were calculated by Muzykantskii and
Khmelnitskii~\cite{Muzykantskii95}, and by Fal'ko and Efetov
~\cite{Falko95}.

The ratio of $M_{pq}$ and its universal ($g \to \infty$) value
$M_{pq}^{(un)}$ can be written ~\cite{to be} as the functional
integral
\begin{eqnarray}
{M_{pq} \over M_{pq}^{(un)}}={1 \over \Xi}\int{\cal D}\theta ({\bf r})e^{-F +
  p(\theta_{1} - \Omega) + q(\theta_{2} - \Omega)},
\label{moments}
\end{eqnarray}
where $\theta_{1,2} = \theta({\bf r}_{1,2})$ and $\Xi = \int {\cal
 D}\theta ({\bf r}) e^{-F}$. Note that while $M^{un}_{pq}$ is
 determined by the zero mode of the diffusion operator, the ratio
 Eq. (\ref{moments}) does not include the integration over this
 mode. As a result the functional integral in Eq. (\ref{moments}) is
 over all functions $\theta ({\bf r})$ that satisfy the condition $\int
 d{\bf r} \theta = 0$, and
\begin{eqnarray}
F[\theta] = {{\pi\beta D} \over {4\Delta}} \int (\nabla
\theta)^2{d{\bf r} \over V},
~~~~\Omega[\theta] = \ln{\left( \int e^{\theta}{d{\bf r} \over V}\right)}.
\label{f,N}
\end{eqnarray}
Eq.(\ref {moments}) enables us to express the two-points mutual distribution
function $P(V|\psi_\alpha ({\bf r_1})|^2,V|\psi_\alpha ({\bf r_2})|^2)$ at
finite $g$ through the universal one $P_{un}(v_1,v_2)$
\begin{eqnarray}
  P(v_1,v_2) = {1 \over \Xi} \int {\cal D} \theta ({\bf r})  e^{-F+2\Omega
-\theta_1 -\theta_2}\nonumber\\ \times P_{un}(v_1e^{\Omega-\theta_1},v_2
e^{\Omega-\theta_2}).
\label{fi:Pv1v2}
\end{eqnarray}

The universal two-point distribution functions $P_{un}(v_1,v_2)$ for
unitary and orthogonal symmetries were determined earlier
~\cite{Prigodin96,Srednicki}:
\begin{eqnarray}
  P^{(u)}_{un}(v_1,v_2)= {\exp{\left(-{v_1+v_2 \over 1-f^2}\right)} \over
    1-f^2} I_0\left({2f\sqrt{v_1v_2}\over 1-f^2}\right),\nonumber\\ 
  P^{(o)}_{un}(v_1,v_2) = {\exp\left[ -{v_1+v_2 \over 2(1-f^2) }\right]\over
    2\pi \sqrt{(1- f^2)v_1 v_2}} \cosh\left({f\sqrt{v_1v_2} \over 1
    -f^2}\right), \nonumber
\end{eqnarray}
\noindent where $f$ is the Friedel function of the distance $r = |{\bf
r}_1 - {\bf r}_{2}|$
\begin{eqnarray}
f(r)=\Gamma(d/2)(2/kr)^{d/2-1}~J_{d/2-1}(kr)e^{-r/(2l)}.\nonumber
\end{eqnarray}
Above $J_p(x)$ and $I_0(x)$ are Bessel and modified Bessel functions,
respectively, $k$ is the wave number. Note that $f(r) \to 0$, when $kr \gg 1$, and
$f(0)=1$.

A usual way to calculate the functional integral like
Eq. (\ref{fi:Pv1v2}) is  to present $\theta ({\bf r})$
as a sum $\sum_\mu \theta_\mu \phi_\mu ({\bf r})$ (there is no
contribution with $\mu = 0$ since $\int d{\bf r}\theta = 0$) over
the eigenfunctions of the problem Eq.(\ref {problem}) thus
reducing the calculation of the functional integral to a sequence
of definite integrals over $\theta_\mu$.

Expansion of the exponent in Eq. (\ref{fi:Pv1v2}) up to the second
order in $\theta_\mu$ leads to a gaussian integral that can be
evaluated explicitly
\begin{eqnarray}
P(v_{1},v_{2}) = \int^{\infty}_{0}\int^{\infty}_0
\frac{ds_{1}ds_{2}}{2\pi} \frac {P_{un} (v_{1}s_{1}, v_{2}
s_{2})}{\sqrt {\Pi_{11} \Pi_{22} - \Pi^{2}_{12}}}\nonumber \\ \times \exp{
\left[ - \sum_{\kappa = \pm1} \frac {(\sqrt{\Pi_{11}} \ln s_{2} +
\kappa \sqrt{\Pi_{22}} \ln s_{1})^{2}}{\Pi_{11} \Pi_{22} +\kappa \Pi_{12}
\sqrt{\Pi_{11} \Pi_{22}}}\right]},
\label{result:Pv1v2}
\end{eqnarray}
\noindent
where $\Pi_{ij}$ is the Green function of  Eq.(\ref{problem})
\begin{equation}
\Pi_{ij} = \frac{2\Delta}{\pi \beta} \sum_{\mu \neq 0}
\frac{\phi_{\mu} ({\bf r}_{i}) \phi_\mu ({\bf r}_j)}{\omega_{\mu}}.
\label{diff.mode}
\end{equation}
For ${\bf r}_1 \to {\bf r}_2$ Eq. (\ref{result:Pv1v2}) reproduces the
result for one-point fluctuations~\cite{Falko95}.

From Eq.(\ref{fi:Pv1v2}) we can guess the probability
density of a given realization $\psi ({\bf r})$ of an
eigenfunction in the form of a functional integral.  Instead of
the gaussian distribution that is valid in the universal
limit~\cite{Alhassid,Srednicki}  we obtain
\begin{eqnarray}
  P[\psi] {\cal D} \psi = {\cal D} \psi {1 \over \Xi}\int {\cal
  D}\theta ({\bf r})e^{-F} \left(\beta^{-1}\det {\hat
  {K}}\right)^{-\beta/2}\nonumber\\ \times \exp{\left[-{\beta\over 2}
  \int d{\bf r}_1\int d{\bf r}_2\psi^*({\bf r}_1){\hat {K}}^{-1}\psi
  ({\bf r}_2)\right]}.
\label{Ppsi}
\end{eqnarray}
The matrix elements of the operator ${\hat {K}}$ in the coordinate
representation $K_{12} \equiv <{\bf r}_{1} \vert\hat { K} \vert {\bf
r}_{2}>$ equal to
\begin{eqnarray}
K_{12} = f(|{\bf r}_1 -{\bf r}_2|)
~\exp \left[ \frac{\theta_1+\theta_2}{2} -
  \Omega \right].
\label{K}
\end{eqnarray}

The distribution determined by Eqs.(\ref{Ppsi},\ref{K}) enables
straightforward calculation of correlation functions for wave function
intensity at many points separated by distance larger than the mean free
path:
\begin{eqnarray}
< \prod_{i} V^{n_i}\vert\psi ({\bf r}_{i}) \vert^{2n_{i}}>={1 \over
\Xi} \int D \theta ({\bf r}) e^{-F}\prod_{i} b_{n_i} K^{n_i}_{ii}.
\label{m.p.c.f.}
\end{eqnarray}
For two points correlation ($i=1,2$)  Eq.(\ref{m.p.c.f.}) directly
follows from Eq. (\ref{moments}). The many points correlation functions
also can be shown to coincide with Eq. (\ref{m.p.c.f.}).

Eq.(\ref{Ppsi}) enables us to present ${\tilde
{P}}_n(s)$ in the form
\begin{eqnarray}
{\tilde {P}}_n(s) = \int {{\cal D}\theta({\bf r})\over \Xi} e^{-F}
\exp\left[-{s\over V} \int (e^{n(\theta-\Omega)}-1)d{\bf r}\right].
\label{func:Ps}
\end{eqnarray}
Using in Eq. (\ref{func:Ps}) the harmonic approximation that is valid for
small fluctuations $u \ll 1$, we obtain Eq. (\ref{result:ps}).

At $u \gg 1$ we can apply the method of optimal fluctuation to
 Eq. (\ref{func:Ps}). For a spherical dot of the radius $R$ the
 saddle-point equations can be written as
\begin{eqnarray}
  {\tilde P}_n(s) \approx \exp{\left\{s -gd\int_0^1\left[\dot
    \theta_c^2 + {\Phi (\theta_c) \over n-1}\right]t^{d-1}dt\right\}},
\label{asym}
\end{eqnarray}
where $t =r/R$, a dot stands fot $t$-derivatives, and $\theta_c (t)$ obeys
the equation
\begin{eqnarray}
  \ddot \theta_c + { d-1 \over t}\dot \theta_c = -{1\over 2}{\delta \Phi \over
    \delta \theta_c},~~\Phi (\theta) = p(e^{n\theta} -ne^\theta),
\label{s.p.}
\end{eqnarray}
supplied by the condition $\dot \theta_c(0) = \dot \theta_c (1) = 0$.  $p$ in
Eq. ({\ref{s.p.}) should be determined from the self-consistency condition
\begin{eqnarray}
p = {d \over 1-n}\left({pg \over -s}\right)^{1/n}\int_0^1\Phi
(\theta_c) t^{d-1}dt.
\nonumber
\end{eqnarray}

Let us note that the potential $\Phi (\theta)$ has a minimum at
$\theta = 0$. At $u \gg 1$ or $\vert s \vert \ll g$ the essential
contribution to Eq.  (\ref{asym}) is given by the trajectories of
$\theta_c(t)$, that starts with $\theta_c(0) \sim -{2\over
n-2}\ln{\vert g/s\vert } \to -\infty $ and ends somewhere near the
minimum of potential $\Phi (\theta)$.  They correspond to the optimal
wave functions that in the center of dot is small like $\vert s/g\vert
^{2/(n-2)}$ and increases when approaching the boundary. Such a
trajectory leads to the asymptotic Eq.(\ref{result:s.p.}) for the
distribution function $P_n(u)$.

Fal'ko and Efetov~\cite{FE} found that there are long range spatial
correlations of eigenfunctions even at $g=\infty$, provided in a
crossover from orthogonal to unitary symmetries, rather than any of
the pure symmetry cases.  Here we have not discussed the crossover
regime.  Calculations similar to ours for the
distributions of moments Eq.(\ref{moments}) and the two point
distribution function Eq.(\ref{result:Pv1v2}) in the crossover regime
can be carried out rather straightforwardly.  Details of this
calculation will be published elsewhere~\cite{to be}.  The statistics
for the crossover regime are determined by so-called ``Cooperon gap''
$\omega_c \approx (\Phi/\Phi_0)^2\omega_1$, where $\Phi$ is the
magnetic flux through the system, and $\Phi_0$ is the flux quantum.
If $\omega_1 \to \infty$ and $\omega_c $ is finite, these results are
in agreement with Ref.~\cite{FE}. In the crossover regime the diffuson
eigenfunctions are not orthogonal to cooperon ones.  Therefore the
distributions of the moments Eq.(\ref{moments}) acquire dependence on
the eigenfunctions.

To summarize, we have calculated the distributions for generalized moments of
inverse participation ratio Eq.(1) in an ensemble of disordered metallic
grains with a given dimensionless conductance $g$. In the universal limit $g
\to \infty$ these moments do not fluctuate due to self-averaging and thus have
definite values Eq.(2). The fluctuations appear only at finite $g$ together
with long-range correlations in densities of wave functions. Indeed, in the
universal limit, when only short-range correlations persist, the fluctuations
vanish when the grain volume (the rank of random matrices) tends to infinity.
Contrarily, at finite $g$ they appear small (as $1/g$) but long range
correlations (see Eq. (\ref{diff.mode})) controlled entirely by the
eigenfunctions of the diffusion operator Eq.(5). These correlations give rise
to the fluctuations of the moments.

It is amazing however that at $g \gg 1$ and for $n \ll \sqrt{g}$ the main body
of the distribution functions $P_n(u)$ (Eqs.
(\ref{meanw},\ref{logzeta},\ref{def:C}-\ref{morew})) can be expressed through
the spectrum of the diffusion operator $\omega_\mu$ and do not depend on the
eigenfunctions $\phi_\mu$. This fact suggests generalization of these
statistics of eigenfunctions from disordered to generic chaotic systems, by
making use of the Ruelle (dynamical) zeta-function (Eqs.
\ref{result:ps},\ref{zeta}), in a way similar to what was done in
Ref.~\cite{AAA} for spectral statistics. Substitution of the spectral
determinant of the diffusion operator by the Ruelle zeta-function for spectral
statistics in generic case was supported by the calculations in the framework
of a nonlinear $\sigma$-model~\cite{AASA}. The analysis of Bogomolny and
Keating~\cite{BK} based on the periodic orbit theory led to similar but
different results. We hope that further analytical development of results on
the eigenfunction statistics together with numerical evaluations will clarify
the relation between quantum statistics and classical dynamics including the
phenomenon of scarring~\cite{scaring}.

The authors are grateful to O. Agam, I. Aleiner, A. Andreev, and K. Efetov for
useful discussions and thank P. Fulde for his interest to the work and
support.


\end{document}